\documentclass[11pt]{book}

\usepackage{amssymb}
\usepackage{amsmath}
\usepackage{amsfonts}
\usepackage{epsfig}

\makeatletter

\newcommand{\newc}{\newcommand}
\newc{\beq}{\begin{equation}}
\newc{\eeq}{\end{equation}}
\newc{\kt}{\rangle}
\newc{\br}{\langle}
\newc{\beqa}{\begin{eqnarray}}
\newc{\eeqa}{\end{eqnarray}}
\newc{\pr}{\prime}
\newc{\longra}{\longrightarrow}
\newc{\ot}{\otimes}
\newc{\rarrow}{\rightarrow}
\newc{\h}{\hat}
\newc{\bom}{\boldmath}
\newc{\btd}{\bigtriangledown}
\newc{\al}{\alpha}
\newc{\be}{\beta}
\newc{\ld}{\lambda}
\newc{\ldmin}{\lambda_{\rm min}}
\newc{\sg}{\sigma}
\newc{\p}{\psi}
\newc{\eps}{\epsilon}
\newc{\om}{\omega}
\newc{\mb}{\mbox}
\newc{\tm}{\times}
\newc{\hu}{\hat{u}}
\newc{\hv}{\hat{v}}
\newc{\hk}{\hat{K}}
\newc{\ra}{\rightarrow}
\newc{\non}{\nonumber}
\newc{\ul}{\underline}
\newc{\hs}{\hspace}
\newc{\longla}{\longleftarrow}
\newc{\ts}{\textstyle}
\newc{\f}{\frac}
\newc{\df}{\dfrac}
\newc{\ovl}{\overline}
\newc{\bc}{\begin{center}}
\newc{\ec}{\end{center}}
\newc{\dg}{\dagger}
\newc{\prh}{\mbox{PR}_H}
\newc{\prq}{\mbox{PR}_q}

\renewenvironment{thebibliography}[1]
     {\section*{\bibname}%
      \@mkboth{\MakeUppercase\bibname}{\MakeUppercase\bibname}%
      \list{\@biblabel{\@arabic\c@enumiv}}%
           {\settowidth\labelwidth{\@biblabel{#1}}%
            \leftmargin\labelwidth
            \advance\leftmargin\labelsep
            \@openbib@code
            \usecounter{enumiv}%
            \let\p@enumiv\@empty
            \renewcommand\theenumiv{\@arabic\c@enumiv}}%
      \sloppy
      \clubpenalty4000
      \@clubpenalty \clubpenalty
      \widowpenalty4000%
      \sfcode`\.\@m}
     {\def\@noitemerr
       {\@latex@warning{Empty `thebibliography' environment}}%
      \endlist}

\makeatother

\newcommand{\sect}[1]{\setcounter{equation}{0}\section{#1}}

\renewcommand\bibname{References}

\begin{document}

\setlength{\baselineskip}{5.0mm}


\chapter[Extreme Eigenvalues of Wishart Matrices]{Extreme Eigenvalues of Wishart Matrices:
Application to Entangled Bipartite System}
\thispagestyle{empty}

\ \\

\noindent
{{\sc Satya N. Majumdar} 
\\~\\
Laboratoire de Physique Th\'{e}orique et Mod\`{e}les
Statistiques (UMR 8626 du CNRS)
Universit\'e Paris-Sud, B\^atiment 100
91405 Orsay Cedex, France}

\begin{center}
{\bf Abstract}
\end{center}
This chapter discusses an application of the random matrix theory
in the context of estimating the bipartite entanglement of a quantum system.
We discuss how the Wishart ensemble (the earliest studied random matrix ensemble)
appears in this quantum problem. 
The eigenvalues of the reduced density matrix of one of the 
subsystems have similar statistical properties as those of the Wishart 
matrices, except
that their {\em trace is constrained to be unity}.
We focus here on the smallest eigenvalue which serves as an important
measure of entanglement between the two subsystems. 
In the hard edge case (when the two subsystems have equal sizes)
one can fully characterize the probability distribution
of the minimum eigenvalue 
for real, complex and quaternion matrices of all sizes. 
In particular, we discuss the important finite size effect due to the {\em fixed trace 
constraint}. 

\sect{Introduction}\label{intro}

The different chapters of this book have already illustrated numerous 
applications
of random matrices in a variety of problems ranging from physics to
finance. In this chapter, I will demonstrate yet another beautiful application
of random matrix theory in a bipartite quantum system that is {\it entangled}.
Entanglement has off late become a rather fashionable subject due to its
applications in quantum information theory and quantum computation.
Entanglement serves as a simple 
measure of nonclassical correlations between different 
parts of a quantum system. The more the entanglement between
two parts of a system, better it is for the functioning of algorithms 
of quantum computation. This is so because, intuitively speaking, quantum 
states 
that are highly entangled contain more informations about different subparts
of the composite system. In this chapter I will discuss the 
statistical properties of the entanglement in a particularly simple
model of the `random pure state' of a bipartite system. We will
see how random matrices come into play in such a system.

Indeed, historically the earliest studied ensemble of random matrices is
the Wishart ensemble (introduced by Wishart~\cite{Wishart} in 1928
in the context of multivariate data analysis, much before 
Wigner introduced the standard Gaussian ensembles of random matrices
in the physics literature). Wishart matrices have found wide applications 
in a variety of systems (see the discussion later and also
chapter 28 and chapter 40 of this book).
In this chapter, we will see that the Wishart ensemble (
with a fixed trace constraint) also appears quite 
naturally 
as the reduced density matrix of a coupled entangled bipartite quantum system.
The plan of this chapter, after a brief introduction to Wishart
matrices, is to explore this connection more deeply with a
particular focus on the statistics of the minimum eigenvalue
which serves as a useful measure of entanglement.
 
Let us start with a brief recollection of the Wishart matrices. Consider a square $(N\times N)$ matrix 
$W$ of the product form $W=XX^{\dagger}$ where $X$ is a $(N\times M)$ rectangular matrix with real, 
complex or quaternion entries and $X^{\dagger}$ its conjugate. The matrix $W$ has a 
simple and natural interpretation. For example, let the entries $X_{ij}$ of the $X$ matrix represent some 
data, e.g., the price of the $i$-th commodity on, say, the $j$-th day of observation. So, 
there are $N$
commodities and for each of them we have the prices for $M$ consecutive days, represented by 
the $(N\times M)$ array $X$. Thus for each commodity, we have $M$ different samples. The product 
$(N\times 
N)$ matrix $W=XX^{\dagger}$ then represents the (unnormalized) covariance matrix, i.e., the correlation 
matrix between the prices of $N$ commodities. If the entries of $X$ are independent Gassian random 
variables chosen
from the joint distribution 
$P[\{X_{ij}\}]\propto \exp\left[-\frac{\beta}{2} {\rm 
Tr}(XX^{\dagger})\right]$ 
(where the Dyson index $\beta=1$, $2$, or $4$ corresponds  
respectively to real, complex or quaternion entries), 
then the random covariance matrix $W$ is called the 
Wishart matrix~\cite{Wishart}.
This ensemble is also referred to as the Laguerre ensemble
since its spectral properties involve Laguerre 
polynomials~\cite{Bronk,Forrester1}.

As mentioned earlier, since its introduction Wishart matrices have found an 
impressive list of applications. They play an 
important role in statistical data analysis~\cite{Wilks,Johnstone}, 
in particular in data compression 
techniques known as Prinicipal Component Analysis (PCA) (see chapter 28
and chapter 40 of this book) with 
applications in
image processing~\cite{Fukunaga}, biological microarrays~\cite{Holter,Alter}, population
genetics~\cite{Patterson,Novembre}, finance~\cite{Bouchaud,Burda}, meteorology
and oceanography~\cite{Preisen} amongst others. In physics, 
Wishart matrices have appeared in multiple 
areas: in nuclear physics~\cite{Fyo1}, in low energy QCD and gauge 
theories ~\cite{QCD} (see also chapter 32 of this book), quantum 
gravity~\cite{QG,Akemann1} and also in several problems in statistical physics. 
These
include directed polymers in a disordered medium~\cite{Johansson1}, nonintersecting
Brownian excursions~\cite{Katori1,Schehr} and 
fluctuating nonintersecting interfaces
over a solid substrate~\cite{Nadal}. Several deformations of Wishart ensemble, with
multiple applications, have also been studied in the literature~\cite{Akemann2}.

The Wishart matrix $W$ has $N$ non-negative random eigenvalues denoted by
$\{w_1,w_2,\ldots,w_N\}$ ($w_i\ge 0$ for each $i$) whose spectral properties  
are well understood and some of them will be briefly reviewed
in section 1.2. These include the joint distribution of $N$ eigenvalues,
the average density of eigenvalues and also the distribution of extreme
eigenvalues (the largest and the smallest). In
this chapter we will be mostly concerned with the distribution of the smallest 
eigenvalue
$w_{\rm min}= {\rm min}(w_1,w_2,\ldots, w_N)$ in the particular case $M=N$  
corresponding to the so called {\em hard edge} (at the origin) case where the average
$\langle w_{\rm min}\rangle \to 0$ as $N\to \infty$. In this case, the properties
of the small eigenvalues (near $w=0$) are governed by Bessel functions
in the large $N$ limit~\cite{Edelman,Forrester1,Nagao1,Verb1,Nagao2}. Such hard edge
properties are absent in the traditional Wigner-Dyson Gaussian 
random matrix~\cite{Mehta}
whose eigenvalues can be both positive and negative. 

The reason we are interested in the smallest eigenvalue distribution 
of the Wishart matrix is because of its application in the seemingly unrelated 
quantum entanglement problem
which is the main objective of this chapter.
As we will see later, Wishart matrices 
will appear naturally as the reduced density matrix in a coupled bipartite 
quantum system 
that is in an {\em entangled random pure state}. 
There is a slight twist though: the Wishart matrix 
in this system satisfies a constraint, namely its {\em trace 
is fixed to unity}.
This {\em fixed trace ensemble} is thus
analogus to the {\em microcanonical} ensemble in statistical mechanics while the 
standard (unconstrained) Wishart ensemble being the analogue of the {\em canonical} ensemble
in statistical mechanics (for other discussions on fixed trace ensembles
see chapter 14, section 14.3.2 of this book). 
In particular, our emphasis will be on the distribution of the smallest
eigenvalue $\lambda_{\rm min}$ in this fixed trace Wishart ensemble.
This is because the smallest eigenvalue turns to be a very useful
observable in this system which contains informations about
entanglement. 
For the special case $M=N$ (hard edge),
we will see that the distribution of 
$\lambda_{\rm min}$ can be exactly computed {\em for all} 
$N$ in this 
fixed trace ensemble in all three cases 
$\beta=1$, $\beta=2$ and $\beta=4$. In particular, we will
discuss how the fixed trace constraint modifies 
the distribution of $\lambda_{\rm min}$ from its 
counterpart in the canonical Wishart ensemble.
We will see that the global
fixed trace constraint 
gives rise to rather strong finite size
effects. This is relevant in the quantum context where
the subsystems can be just a few qubits. So, it is actually
important to know the distribution of entanglement for
finite size systems (the thermodynamic limit is not always
relevant in this context). Hence, the fact that one can
compute the distribution of the minimum eigenvalue exactly
for all $N$ (not necessarily large) in presence of the fixed trace 
constraint becomes important and relevant. 

This rest of the chapter is organized as follows. In section \ref{WishartSP}, 
we briefly review
some spectral properties of unconstrained Wishart matrices. In section \ref{Entang}
we introduce the problem of the random pure state of an entangled
quantum bipartite system. Its connection to Wishart matrices with
a {\rm fixed trace constraint} is established. Next we focus
on the smallest eigenvalue and derive
its probability distribution for the bipartite problem
in section \ref{minimum}. Finally we conclude in section \ref{conclu}
with a summary and open problems.

\sect{Spectral Properties of Wishart Matrices: A brief summary}\label{WishartSP}

Let us first briefly recall some spectral properties of the $(N\times N)$ Wishart 
matrix 
$W=XX^{\dagger}$ with $X$ being a rectangular $(N\times M)$ matrix with real $(\beta=1)$,
complex $(\beta=2)$ or quaternion $(\beta=4)$ Gaussian entries drawn from
the joint distribution 
$P[\{X_{ij}\}]\propto \exp\left[-\frac{\beta}{2} {\rm 
Tr}(XX^{\dagger})\right]$.
These 
results will be useful for the problem of the random pure state 
of the bipartite system to be discussed in the next section.
\\

{\noindent \bf {Joint distribution of eigenvalues:}} The $N$ eigenvalues of $W$, 
denoted by $\{w_1,w_2,\ldots, w_N\}$, are non-negative and have 
the joint probability density function (pdf)~\cite{James}
\begin{equation}
P[\{w_i\}]= K_{N,M}\, e^{-\frac{\beta}{2}\sum_{i=1}^N w_i}\, \prod_{i=1}^N 
w_i^{\alpha\beta/2}\,\prod_{j<k} |w_j-w_k|^{\beta}
\label{jpdwish1}
\end{equation}
where $\alpha= (1+M-N)-2/\beta$ and the normalization constant $K_{N,M}$ can be computed 
exactly~\cite{James}. 
Without any loss of generality, we will assume $N\le M$. This is because if $N>M$,
one can show that $N-M$ eigenvalues are exactly $0$ and the rest of the $M$
eigenvalues are distributed exactly as in Eq. (\ref{jpdwish1}) with $N$ and $M$
exchanged. Note that while for Wishart matrices $M-N$ is a non-negative {\em integer}
and $\beta=1$, $2$ or $4$,
the joint density in Eq. (\ref{jpdwish1}) is well defined for any $\beta>0$ and
$\alpha>-2/\beta$ (this last condition is necessary so that the joint pdf
is normalizable). When 
these parameters take continuous values the
joint pdf is called the Laguerre ensemble. \\

{\noindent \bf {Coulomb gas interpretation, typical scaling and average density of states:}} 
The joint
pdf (\ref{jpdwish1}) can be written in the standard Boltzmann form,
$P[\{w_i\}]\propto \exp\left[-\beta E(\{w_i\})\right]$ where 
\begin{equation}
E[\{w_i\}= \frac{1}{2}\sum_{i=1}^N (w_i-\alpha \log w_i)-\frac{1}{2} \sum_{j\ne 
k} \ln |w_j-w_k|
\label{coulomb1}
\end{equation}
can be identified as the energy of a Coulomb gas of charges with
positions $\{w_i\}\ge 0$. These charges  
repel each other via the $2$-d Coulomb (logarithmic) interaction (the second 
term
in the energy), though
they are restricted to live on the positive real line. 
In addition, these charges are subjected to an external potential
which is linear+logarithmic (the first term in the energy ). 
The external potential
tends to push the charges towards the origin while the Coulomb repulsion tends
to spread them apart. The first term typically scales as $w_{\rm typ} N$ 
where $w_{\rm typ}$ is the typical value of an eigenvalue,
while
the second term scales as $N^2$ for large $N$. Balancing these two terms
one gets $w_{\rm typ} \sim N$ for large $N$. Indeed, this scaling shows
up in the average density of states (average charge density) which can be computed from the
joint pdf and has the following scaling for large $N$
\begin{equation}
\rho_N(w) = \frac{1}{N}\sum_{i=1}^N \langle \delta(w-w_i)\rangle \to \frac{1}{N}\, f_{\rm 
MP}\left(\frac{w}{N}\right)
\label{dens1}
\end{equation}
where the Marcenko-Pastur(MP) scaling function is given by~\cite{MarPas}
(see also  chapter 28 section 28.4.1 of this book)
\begin{equation}
f_{\rm MP}(x) = \frac{1}{2\pi x}\, \sqrt{(b-x)(x-a)}.
\label{MP1}
\end{equation}
Thus the charge density is confined over a finite support $[a,b]$ with
the lower edge 
$a= (1-c^{-1/2})^2$ and the upper edge $b=(1+c^{-1/2})^2$ with $0\le c=N/M \le 1$.
For all $c<1$, the average density vanishes at both edges of the MP sea. For the special
case $c=1$ (this happens in the large $N$ limit when $M-N<< O(N)$), the lower 
edge $a$ gets pushed towards the
hard wall at $0$ (this is the so called {\em hard edge limit}) and the upper edge $b\to 4$
and the average density simply becomes, $f_{\rm MP}(x) =\frac{1}{2\pi} \sqrt{(4-x)/x}$
for $0\le x\le 4$. It diverges as $x^{-1/2}$ at the hard lower edge $x=0$. 

For later purposes, it is also useful to calculate the average value of the
trace $Tr= \sum_{i=1}^N w_i$. Using the expression for the average density of
states, it follows that for large $N$
\begin{equation}
\langle Tr \rangle = N \int_0^{\infty} w\,\rho_N(w)\, dw\to  \frac{N^2}{c}. 
\label{avgtr1}
\end{equation}
In particular, for $c=1$ (i.e., $M-N<< O(N)$), we have 
$\langle Tr \rangle =N^2$ in
the large $N$ limit. \\

{\noindent \bf {Maximum eigenvalue:}} Let
$w_{\rm max}={\rm max}(w_1,w_2,\ldots,w_N)$ denote the
maximum eigenvalue. On an average, it
is located at the upper edge of the MP density of states. It then follows from 
Eq. (\ref{dens1}) that $\langle w_{\rm max}\rangle = b N$ for large $N$.
However, for large but finite $N$, the random variable $w_{\rm max}$ 
fluctuates, from one sample to another, around its average
value $bN$. The typical fluctuations around its mean were shown to be $\sim O(N^{1/3})$
for large $N$~\cite{Johansson1,Johnstone} and the limiting distribution
of these typical fluctuations are described by the well known
Tracy-Widom density~\cite{TW1}. In other words,
$w_{\rm max}= b N + c^{1/6} b^{2/3} N^{1/3} \chi$, where the random variable
$\chi$ has an $N$-independent limiting pdf, $g_{\beta}(\chi)$
described by the Tracy-Widom function~\cite{TW1}.
In contrast, {\em atypically large}, e.g., $\sim O(N)$
fluctuations of $w_{\rm max}$ from its mean are {\em not described} by the   
TW density. Such large fluctuations play an important role
in many practical applications such as in PCA~\cite{VMB,MV}. 
Far away from the mean $bN$, these atypically large fluctuations
of $w_{\rm max}$ are instead described by large deviation functions associated
with the pdf of $P(w_{\rm max},N)$ and are of the form
\begin{eqnarray}
P(w_{\rm max}=t,N) &\sim & \exp\left[-\beta N^2 
\Phi_{-}\left(\frac{bN-t}{N}\right)\right]\quad 
{\rm for}\quad t<< bN \label{lldv} \\
&\sim & \exp\left[-\beta N \Phi_{+}\left(\frac{t-bN}{N}\right)\right]\quad 
{\rm for}\quad t>> bN .\label{rldv}
\end{eqnarray}
The left rate function $\Phi_{-}(x)$ was computed explicitly for all $c$ in~\cite{VMB}
extending a Coulomb gas approach developed originally in~\cite{DM} to compute
the corresponding left rate functions for Wigner-Dyson Gaussian matrices. 
On the other hand, the computation of the right rate function $\Phi_{+}(x)$ 
required a different approach and was recently obtained explicitly for all
$c$~\cite{MV}. The right rate function
in the Wigner-Dyson Gaussian case was also obtained by a different, albeit rigorous, method in 
~\cite{Benarous}. One interesting point is that while the limiting TW density 
$g_{\beta}(\chi)$ depends
on $\beta$, the rate functions $\Phi_{\mp}(x)$ are independent of
$\beta$. \\

{\noindent \bf {Minimum eigenvalue:}} Since in this chapter our main interest in the 
problem 
of
bipartitite entanglement concerns the lowest eigenvalue of the
reduced density matrix, we need to discuss, in some detail,
the statistical properties of the minimum eigenvalue of the
unconstrained Wishart ensemble. For
the minimum 
eigenvalue, 
$w_{\rm min}={\rm 
min}(w_1,w_2,\ldots,w_N)$, the situation is rather different for $c<1$ and $c=1$ cases.
For $c<1$, the lower edge of the MP sea is at $a=(1-c^{-1/2})^2>0$, indicating
that $\langle w_{\rm min}\rangle = a N$ in the large $N$ limit and thus the
typical value of $w_{\rm min}\sim O(N)$. The typical fluctuations of $w_{\rm min}$ around
this mean value $aN$ are again described by the TW density (appropriately rescaled).
The large deviation functions describing atypical fluctuations, to our knowledge, have not 
been systematically
studied as in the maximum eigenvalue case (though see~\cite{Chen} and references
therein). 

The situation, however, is quite different in the $c=1$ case (when
$M-N<< O(N)$ where the
lower edge of the MP sea $a\to 0$. This is the so called {\em hard edge} case. 
We will see shortly that in this case the typical value of the minimum
eigenvalue scales as $w_{\rm min}\sim 1/N$ for large $N$, to be contrasted
with the behavior $w_{\rm min}\sim aN$ for $c<1$. 
There
have been a lot of studies on the distribution of $w_{\rm min}$ in this 
hard edge
$c=1$ ($M-N<< O(N)$) case,  
notably
by Edelman~\cite{Edelman} and Forrester~\cite{Forrester1,
Forrester2}. 
It has also found very nice applications in QCD (see e.g.
chapter 32, section 32.2.6 of this book).
Here, for simplicity, we will focus on the special case
$M=N$ (such that $c=1$ strictly for all $N$, and not just for large $N$).
For other cases when $M-N\sim O(1)$ (so that $c=1$ only in the
large $N$ limit), a summary
can be found in the table 32.2 of chapter 32 of this book (see
also section 4.2 of ~\cite{Akemann2} and references therein).

In this special case $M=N$, the
cumulative distribution of the minimum, $Q_N(z)= {\rm Prob}[w_{\rm min}\ge 
z,N]$,
is known~\cite{Edelman} exactly {\em for all} $N$ in all the three cases $\beta=1$, $\beta=2$ 
and $\beta=4$. Note that, $Q_N(z)=\int_{z}^{\infty}\ldots\int_{z}^{\infty}P[\{w_i\}]\,\prod 
dw_i$ where $P[\{w_i\}]$ is the joint pdf given in Eq. (\ref{jpdwish1}). For $M=N$,
this multiple integral $Q_N(z)$ can be easily performed for $\beta=2$ by making
a trivial shift $w_i\to w_i+z$ and one gets for all $N$
\begin{equation}
Q_N(z)= \exp[-Nz]; \quad \beta=2 
\label{b2}
\end{equation}
For $\beta=1$ and $4$, the simple shift does not work. However, the integral $Q_N(z)$ can
be calculated explicitly~\cite{Edelman}. For $\beta=1$, one obtains for all $N$
\begin{equation}
Q_N(z)= \frac{\Gamma(N+1)}{2^{N-1/2}\Gamma(N/2)}\,\int_z^{\infty}y^{-1/2}\, e^{-Ny/2}\,
U\left(\frac{N-1}{2},-\frac{1}{2},\frac{y}{2}\right)\, dy
\label{b1}
\end{equation}
where $U(p,q,z)$ is the confluent (Kummer) hypergeometric function~\cite{AS}. For $\beta=4$, 
while Edelman
does not provide an explicit expression for $Q_N(z)$, it is not difficult to
obtain $Q_N(z)$ by using his Lemma 9.2~\cite{Edelman} and one gets (see also
~\cite{Forrester2})
\begin{equation}
Q_N(z)=  e^{-2Nz}\, _1F_1\left(-N;\frac{1}{2};-z\right); \quad \beta=4 
\label{b4}
\end{equation}
where 
\begin{equation}
_1F_1(p;q;z)= 1+ \frac{p}{q}\, \frac{z}{1!}+ \frac{p(p+1)}{q(q+1)}\,\frac{z^2}{2!}+\ldots
\label{dhyper}
\end{equation}
is the degenerate hypergeometric function~\cite{AS}. 

The large $N$ limit is interesting where in all three cases $\beta=1$, $2$ and $4$, the
cumulative distribution of the minimum $Q_N(z)$ approaches a scaling form: $Q_N(z)\to 
q_{\beta}(zN)$
where the scaling function $q_{\beta}(y)$ can be computed explicitly
\begin{eqnarray}
q_1(y) &=& \exp\left[-\sqrt{y}-y/2\right] \label{sc1} \\
q_2(y) &=& \exp [-y] \label{sc2} \\
q_4(y) &=& \frac{1}{2}\,\left[e^{-2y+2\sqrt{y}}+e^{-2y-2\sqrt{y}}\right]. \label{sc4} 
\end{eqnarray}
Note in particular that the average value $\langle w_{\rm min} \rangle =\int_0^{\infty} 
Q_N(z) dz \to c_\beta/N$
for large $N$, where the prefactor 
\begin{equation}
c_\beta= \int_0^{\infty}q_{\beta}(y) dy
\label{prefac1}
\end{equation}
can be computed explicitly in all three cases and one gets
\begin{eqnarray}
c_1 &=& 2\left[1- \sqrt{\frac{\pi e}{2}}\, {\rm erfc}(1/\sqrt{2})\right]=0.68864.. 
\label{c1} \\
c_2 &=& 1 \label{c2} \\
c_4 &=& \frac{1}{2}\,\left[1+\sqrt{\frac{\pi e}{2}}\,{\rm 
erf}(1/\sqrt{2})\right]=1.20534.. \label{c4} 
\end{eqnarray}
where ${\rm erf}(z)= \frac{2}{\sqrt{\pi}}\int_0^{z} e^{-u^2}\, du$ is the
standard error function and ${\rm erfc}(z)= 1-{\rm erf}(z)$. These results will
be used in Section \ref{minimum}.

\sect{Entangled Random Pure State of a Bipartite System}\label{Entang}

We now turn to the main problem of interest in this chapter, namely the
properties of an entangled random state of a quantum bipartite system.
We will see that Wishart matrices, albeit with a {\em fixed trace constraint},
play a central role in this problem. 

As mentioned in the introduction, entanglement has been studied extensively in the 
recent past 
due to its central role in
quantum information and possible involvement in quantum computation. 
In the context of quantum algorithms, it is often desirable to create 
states of large
entanglement. 
A potential candidate for such a state with `large entanglement'
that is relatively simple to analyse 
turns out to be the `random pure state' in a bipartite system~\cite{Hayden}
which we will describe in detail shortly.
Such a random pure state can also be used as a null model or reference point
to which the entanglement of an arbitrary time-evolving state may 
be compared.
Apart from the issue of bipartite entanglement, 
statistical properties of such random states
are relevant for quantum chaotic or non-integrable systems.
The applicability of random matrix theory and hence of random states 
to systems with 
well-defined
chaotic classical limits was pointed out
long back~\cite{Bohigas84}. 

We start with a general discussion of entanglement 
in a bipartite setting without any reference to any specific statistical
measure. The statistical properties will be discussed later when
we introduce the `random' state. For now, the discussion below
holds for any quantum pure state.
Let us consider a composite
bipartite system $A\otimes B$ composed of two smaller subsystems $A$ and $B$, whose
respective Hilbert spaces ${\cal H}^{(N)}_A$ and ${\cal H}^{(M)}_B$ have dimensions
$N$ and $M$. The Hilbert space of the composite system ${\cal
H}^{(NM)}={\cal H}^{(N)}_A \otimes {\cal H}^{(M)}_B$ is thus $NM$-dimensional.
Without
loss of generality we will assume that $N\le M$. Let $\{|i^A\kt \}$ and
$\{|\alpha^B \kt \}$ represent two complete basis states for $A$ and $B$
respectively. Then, any arbitrary pure state $|\psi \kt$ of the composite system
can be most generally written as a linear combination
\beq
|\psi\kt = \sum_{i=1}^N\sum_{\alpha=1}^M x_{i,\alpha}\, |i^A\kt\otimes |\alpha^B\kt
\label{state1}
\eeq
where the coefficients $x_{i,\alpha}$'s form the entries of a rectangular $(N\times M)$ matrix
$X=[x_{i,\alpha}]$.
As an example of such a bipartite system, $A$ may be considered a given
subsystem (say a set of spins) and $B$ may represent the environment (e.g., a
heat bath). 

Next we discuss the density matrix and the concept of entanglement.
For a pure state, the density matrix of the composite system
is simply defined as
\beq
\rho = |\psi\kt\, \br\psi|
\label{dm1}
\eeq
with the constraint ${\rm Tr}[\rho]=1$, or equivalently $\br 
\psi|\psi\kt=1$. Note that had the composite system been in a statistically {\it mixed} state, 
its density
matrix would have been of the form
\beq
\rho = \sum_{k} p_k\, |\psi_k\kt\, \br \psi_k|,
\label{dmmixed}
\eeq
where $|\psi_k\kt$'s are the pure states of the composite system and $0\le p_k\le 1$
denotes the probability that the composite system is in the $k$-th pure state, with
$\sum_k p_k=1$. A classical example of such a mixed state is when the
system is in the canonical ensemble at given temperature $T$: in this case
the density matrix is given by
\beq
\rho= \sum_E \frac{1}{Z} e^{-E/{k_B T}}\, |E\kt\, \br E|
\label{can1}
\eeq
where $Z=\sum_E e^{-E/{k_B T}}$ is the canonical partition function
($k_B$ is the Boltzmann constant) and the pure state $E\kt$ denotes the energy
eigenstate (with eigenvalue $E$) of the full system. 
In this chapter, we will not discuss the mixed state and will 
restrict ourselves only to the case of a {\em pure} state whose
density matrix is given in Eq.~(\ref{dm1}).

The concept of entanglement is simple. A pure state $|\psi\kt $ is
called {\bf entangled} if it is {\it not} expressible as a direct
product
of two states belonging to the two subsystems $A$ and $B$.
Only in the special case when the coefficients have the product form,
$x_{i,\alpha}= a_i b_{\alpha}$ for all $i$ and $\alpha$, the state $|\psi \kt=
|\phi^A\kt \otimes |\phi^B \kt $ can be written as a direct product of two states
$|\phi^A \kt=\sum_{i=1}^N a_i |i^A\kt$ and
$|\phi^B\kt= \sum_{\alpha=1}^M b_\alpha |\alpha^B\kt$
belonging respectively to the two subsystems $A$ and $B$. In this case, the
composite state $|\psi \kt$ is fully {\it separable} or {\it unentangled}. But 
otherwise, it is
generically {\it entangled}. \\

Upon using the decomposition in Eq.~(\ref{state1}), the density
matrix of the pure state can be expressed 
as
\beq
\rho = \sum_{i,\alpha}\sum_{j,\beta} x_{i,\alpha}\, x_{j,\beta}^*\, |i^A\kt\br j^A|\otimes
|\alpha^B\kt
\br\beta^B|
\label{dem2}
\eeq
where the Roman indices $i$ and $j$ run from $1$ to $N$ and the Greek indices $\alpha$ and
$\beta$ run from $1$ to $M$. We also assume that the pure state $|\psi\kt$ is normalized to
unity so that ${\rm Tr}[\rho]=1$. 
Hence the coefficients $x_{i,\alpha}$'s must satisfy $\sum_{i,\alpha} 
|x_{i,\alpha}|^2=1$.

Given the density matrix of the pure composite state in Eq. (\ref{dem2}), one can
then compute the reduced density matrix of, say, the subsystem $A$ by tracing over
the states of the subsystem $B$
\beq
\rho_A = {\rm Tr}_B[\rho]=\sum_{\alpha=1}^M \br \alpha^B|\rho|\alpha^B\kt.
\label{rdm1}
\eeq
The reduced density matrix is important because if we measure any observable
$\hat O$ of the subsystem $A$, its expected value is given by
${\rm Tr}[{\hat O}\,\rho_A]$. Thus, $\rho_A$ is the basic 
physical object whose properties are directly related
to measurements.
  
Using (\ref{dem2}) one gets
\beq
\rho_A = \sum_{i,j=1}^N \sum_{\alpha=1}^M x_{i,\alpha}\, x_{j,\alpha}^*\, |i^A\kt\br
j^A|=\sum_{i,j=1}^N W_{ij} |i^A\kt\br j^A|
\label{rdm2}
\eeq
where $W_{ij}$'s are the entries of the $N\times N$ square matrix $W=X X^{\dagger}$.
In a similar way, one can express the reduced density matrix $\rho_B={\rm Tr}_A[\rho]$ of the
subsystem $B$
in terms of the square $M\times M$ dimensional matrix ${\tilde W}=X^{\dagger}X$.
Hence we see how the Wishart covariance matrix $W= XX^{\dagger}$
appears in this quantum problem.
 
Let $\lambda_1,\lambda_2,\ldots,\lambda_N$ denote the $N$ eigenvalues of $W=XX^{\dagger}$.
Note that these eigenvalues are non-negative, $\lambda_i\ge 0$ for all $i=1,2,\ldots,N$.
Now the matrix ${\tilde W}=X^{\dagger} X$ has $M\ge N$ eigenvalues. It is easy 
to prove that 
$M-N$ of 
them
are identically $0$ and $N$ nonzero eigenvalues of ${\tilde W}$ are 
the same as
those of $W$. Thus, in this diagonal representation, one can express $\rho_A$ as
\beq
\rho_A= \sum_{i=1}^N \lambda_i \, |\ld^A_i\kt\, \br \ld^A_i|
\label{diagA}
\eeq
where $|\ld^A_i \kt$'s are the eigenvectors of $W=XX^{\dagger}$. A similar
representation holds for $\rho_B$. It then follows that one can represent
the original composite state $|\psi\kt$ in this diagonal representation as
\beq
|\psi\kt = \sum_{i=1}^{N} \sqrt{\ld_i}\,\, |\ld_i^A\kt \otimes |\ld^B_i \kt
\label{Sch1}
\eeq
where $|\ld^A_i \kt$ and $|\ld^B_i \kt$ represent the normalized 
eigenvectors (corresponding to the same nonzero eigenvalue $\lambda_i$) of 
$W=XX^{\dagger}$ and ${\tilde W}=X^{\dagger} X$ respectively.
This spectral decomposition in Eq. (\ref{Sch1}) is known as the Schimdt decomposition. The
normalization
condition $\br \psi|\psi\kt=1$, or equivalently ${\rm Tr}[\rho]=1$, thus 
imposes
a constraint on the eigenvalues, $\sum_{i=1}^N \ld_i=1$.

Note that while each individual state $|\ld_i^A\kt \otimes|\ld^B_i \kt$ in the Schimdt 
decomposition
in Eq. (\ref{Sch1}) is {\it separable}, their linear combination $|\psi\kt$, in 
general,
is {\it entangled}. This simply means that the composite state $|\psi\kt$ can not, in general,
be written as a direct product $|\psi\kt= |\phi^A\kt \otimes |\phi^B\kt$ of two states of the
respective subsystems. The spectral properties of the matrix $W$, i.e., the knowledge
of the eigenvalues $\lambda_1,\lambda_2,\ldots, \lambda_N$, in association
with the Schimdt decomposition in Eq. (\ref{Sch1}), provide useful information
about how entangled a pure state is.\\

{\noindent {\bf Measures of Entanglement:}} It is useful to construct
a measure of entanglement, i.e., a function of the eigenvalues $\lambda_i$'s
whose value will tell us how entangled a pure state is. There are many ways
of constructing such a measure. Its value should monotonically increase
from the configuration of $\lambda_i$'s where the state is fully
{\em separable} to the configuration where the state is {\em maximally}
entangled. These two configurations, recalling that $\sum_i\lambda_i=1$,
are the following:\\

(i) {\em separable}: When one of the eigenvalues, say $\lambda_1$ is $1$ 
and
the rest are all identically zero. Then the state completely decouples as only
one term, say the first term, is present in Eq. (\ref{Sch1}).\\

(ii) {\em maximally entangled}: When all eigenvalues are equal, i.e., $\lambda_i=1/N$.
In this case all $N$ terms in Eq. (\ref{Sch1}) are present.\\

\begin{figure}[htb]
\begin{center}
\includegraphics[width=8cm]{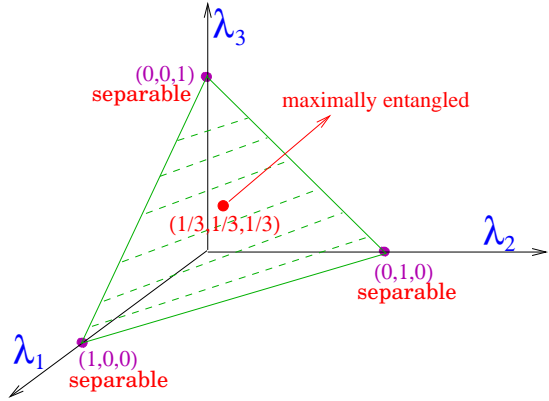}
\caption{A cartoon for $N=3$. The system lives on the triangular
plane $\lambda_1+\lambda_2+\lambda_3=1$. The vertices of the
triangle represent the {\em separable} configurations 
and the centroid $(\lambda_1=\lambda_2=\lambda_3=1/3)$ represents
the {\em maximally entangled} configuration.
\label{fig:triangle}}
\end{center}
\end{figure}

In Fig. (\ref{fig:triangle}), we present a cartoon for $N=3$ for 
the purpose of illustration.
In the three dimensional space $(\lambda_1, \lambda_2,\lambda_3)$,
any point on the triangular plane $\lambda_1+\lambda_2+\lambda_3=1$
with $\lambda_i\ge 0$ represents an allowed configuration. The
three vertices, where the system gets completely factorised, represent
the fully {\it separable} configurations (situation (i) above). On the other 
hand, 
the centroid
$(1/3,1/3,1/3)$ represents the fully (maximally) entangled configuration
(situation (ii) above).

If the system is in a given configuration $\{\lambda_i\}$ on the 
allowed plane $\sum_{i=1}^N \lambda_i=1$ (and $\lambda_i\ge 0$), how 
much entangled 
it is?
In other words, how do we measure the entanglement content 
of a given configuration of $\{\lambda_i\}$? This is
usually done by defining the so called entanglement entropy, a
single scalar quantity associated with each configuration, i.e., each
point on the plane
$\sum_{i=1}^N \lambda_i=1$ (and $\lambda_i\ge 0$). 
A standard and perhaps most studied measure of entanglement is the so called
von Neumann entropy~\cite{Bennet}, $S_1=  -\sum_{i=1}^N \ld_i \ln 
(\ld_i)=-{\rm Tr}[\rho_A \ln(\rho_A)]$, which
has its smallest value $S_1=0$ in configuration (i) and its maximum possible
value $S_1= \ln N$ in configuration (ii). Renyi entropy defined as
$S_q=\ln\left(\sum_{i} \lambda_i^q\right)/(1-q)$~\cite{Renyi} with
the parameter $q>0$ is a
natural generalization that reduces to the von Neumann entropy when $q\to 1$.
Again, for any $q>0$, $S_q=0$ at the `separable' vertices (situation (i)
above) and $S_q=\ln (N)$ at the `maximally entangled' centroid (situation (ii)
above).
For $q=2$, $\sum_{i=1}^N \lambda_i^2= \exp[-S_2]$ is called the purity
that has been widely studied (see \cite{Facchi0} and references therein). For
other measures we refer the reader to the introduction in \cite{Giraud}.
Essentially, one can define any scalar quantity whose value increases 
monotonically as one moves from fully `separable' to maximally `entangled' 
configurations (e.g., as one moves from the vertices towards the 
centroid in Fig. (\ref{fig:triangle})) (for more detailed prescriptions
and requirements
on the measure, see e.g. ~\cite{Vedral}).

Important informations regarding the degree of entanglement can also
be obtained from the two extreme eigenvalues, the
largest $\lambda_{\rm max}={\rm
max}(\lambda_1,\lambda_2,\ldots, \lambda_N)$ and the smallest
$\lambda_{\rm min}={\rm
min}(\lambda_1,\lambda_2,\ldots, \lambda_N)$. 
Due to the constraint $\sum_{i=1}^N \ld_i=1$ and the fact that
eigenvalues are all non-negative, it follows that $1/N\le \ld_{\rm max}\le 1$
and $0\le \ld_{\rm min} \le 1/N$.
Consider, for
instance, the following
limiting situations. Suppose that the largest eigenvalue $\lambda_{\rm max}={\rm
max}(\lambda_1,\lambda_2,\ldots, \lambda_N)$ takes its maximum allowed value $1$. Then due to
the constraint $\sum_{i=1}^N \lambda_i=1$ and the fact that $\lambda_i\ge 0$ for all $i$,
it follows that all the rest $(N-1)$ eigenvalues must be identically $0$. 
Thus it corresponds to the configuration (i) above of fully separable state.
On the other hand, if $\lambda_{\rm max}=1/N$ (i.e., it takes its lowest allowed value),
it follows that all the eigenvalues must have the same value, $\lambda_i=1/N$ for all $i$,
again due to the constraint $\sum_{i=1}^N \lambda_i=1$. This then corresponds
to situation (ii) of maximally entangled state. Thus, for instance,
one can consider $-\ln(\lambda_{\rm max})$
as a measure of entanglement as it increases from its value $0$ in the
separable state to its maximal
value $\ln (N)$ in the maximally entangled case. In fact, $-\ln(\lambda_{\rm 
max})$ 
is precisely the $q\to \infty$ limit of the Renyi entropy $S_q$.

In this chapter our particular interest is on the smallest eigenvalue $0\le \ldmin\le 
1/N$. 
When $\ldmin$ takes its maximal allowed value $1/N$, it follows
again, from the constraint $\sum_{i=1}^N \lambda_i=1$ and $\lambda_i\ge 0$ for 
all $i$,
that all 
the eigenvalues
must have the same value $\lambda_i=1/N$. This will thus make the state
$|\psi \kt$  {\em maximally} entangled, i.e., situation (ii). In the opposite case, when 
$\ldmin$ takes its 
smallest
allowed value $0$, while it does not provide any information on the entanglement
of the state $|\psi\kt$, one sees from the Schmidt decomposition (\ref{Sch1}) that the
dimension of the effective Hilbert space of the subsystem $A$ reduces
from $N$ to $N-1$ (assuming that $\ldmin$ is non-degenerate). 
Indeed, if $\ldmin$ is very close to zero, one can
effectively ignore the term containing $\ldmin$ in Eq. (\ref{Sch1}) and 
achieve a reduced Hilbert space, a process called `dimensional reduction' that is
often used in the compression of large data structures in computer 
vision~\cite{Wilks,Fukunaga,VMB}.
Thus the knowledge of
$\ldmin$ and in particular its proximity
to its upper and lower limits provide informations on both the degree of entanglement
as well as on the efficiency of the dimensional reduction process.\\

{\noindent {\bf Random Pure State:} So far, our discussion was valid for an arbitrary pure state in Eq. 
(\ref{state1}) with
any fixed coefficient matrix $X=[x_{i,\alpha}]$. 
One can now introduce a statistical measure or distribution
for the entries of the matrix $X$ which, in turn, will
induce a probability distribution for the eigenvalues
$\lambda_i$'s of $W=XX^{\dagger}$  that appear in the Schimdt
representation in Eq. (\ref{Sch1}). As a result, any measure
of entanglement (e.g., the von Neumann entropy or the minimum
eigenvalue $\lambda_{\rm min}$) will also have a statistical
distribution associated with it. The main challenge then
is to compute this probability distribution of the
entanglement, given the measure on the entries of $X$.

So, what is an appropriate measure on $X$? Evidently, we can
not choose any arbitrary measure on $X$. Indeed, in Eq. (\ref{state1})
we can just rotate (unitarily) the basis of the Hilbert space.
Clearly physical properties of the system in a pure state should not depend
on which basis we choose. Thus the joint probability distribution
of the entries $x_{i,\alpha}$ in Eq. (\ref{state1}) should
be invariant under a unitary (or orthogonal if we
restrict $X$ to be real) transformation $|\psi \kt \to U |\psi \kt$, where
$U$ represents a unitary operator.
The only measure that remains invariant under a unitary rotation
is the uniform measure over all pure states. This is the so
called Haar measure where the coefficients 
$x_{i,\alpha}$'s are uniformly distributed over all
possible values satisfying the
constraint $\sum_{i,\alpha} |x_{i,\alpha}|^2=1$ or
equivalently,
$P[\{X_{ij}\}]\propto \delta\left({\rm Tr}(XX^{\dagger}-1\right)$.
The physical meaning of this Haar measure is clear: under unitary
time evolution, and in absence of any other conservation law (such as
fixed energy as in the case of standard microcanonical ensemble in statistical
physics),
the system visits all allowed normalized pure states equally
likely, i.e., Haar measure is a natural stationary measure under
unitary evolution when
ergodicity holds over all allowed pure states of the composite
system. This is, in
fact, the case in many physical situations
when the system is described by a sufficiently complex `time-dependent'
Hamiltonian as in quantum chaotic systems~\cite{Arul1}.

Given that the entries of $X$ are distributed via the Haar measure
$P[\{X_{ij}\}]\propto \delta\left({\rm Tr}(XX^{\dagger}-1\right)$, the
next question is how are the eigenvalues of $W=XX^{\dagger}$ distributed?
Noticing that the eigenvalues $\lambda_i$'s of $W=XX^{\dagger}$ are the same as
Wishart eigenvalues, except with the additional constraint ${\rm 
Tr}(W)=\sum_{i=1}^N 
\lambda_i=1$, it follows immediately from 
Eq. (\ref{jpdwish1}) that the joint pdf of $\lambda_i$'s is given by~\cite{LP,ZS}
\beq
P[\{\ld_i\}]=B_{M,N} \delta\left(\sum_{i=1}^N \ld_i -1 \right)
\prod_{i=1}^{N} \ld_i^{\frac{\beta}{2}(M-N+1)-1} \prod_{j<k} |\ld_j-\ld_k|^\beta
\label{jpdf1}
\eeq
where the normalization constant $B_{M,N}$
is known explicitly~\cite{ZS}. 
Note that the exponential factor $e^{-\frac{\beta}{2}\sum_{i=1}^N \lambda_i}$
present in Eq. (\ref{jpdwish1}) becomes a constant due to the constraint
$\sum_{i=1}^N\lambda_i=1$ and hence is absorbed in the normalization constant
$B_{M,N}$.
The
ensemble described in Eq. (\ref{jpdf1}) can thus be seen as the microcanonical
version of the canonical Wishart ensemble in (\ref{jpdwish1}). 

When the coefficient matrix $X$ is drawn
from the Haar measure, we will refer to the state in Eq. (\ref{state1})
as a {\it random pure} state.
Given that $\lambda_i$'s corresponding to the random pure state
are distributed via the joint pdf (\ref{jpdf1}), it follows that 
the associated observables such as the von Neumann 
entropy $S_1=-\lambda_i \ln
(\lambda_i)$, the maximum eigenvalue $\lambda_{\rm max}$, the minimum
eigenvalue $\lambda_{\rm min}$ etc. are also random variables.
The main technical
problem then is to evaluate the statistical properties (such as the mean, variance
or even the full probability 
distribution) of such observables. 

There have been quite a few studies 
in this direction. For example,
the average entropy $\br S_1\kt $ (where the average is performed with the
measure in Eq. (\ref{jpdf1})) was computed for $\beta=2$ by Page~\cite{Page95} and was
found to be $\br S \kt \approx \ln (N)-\frac{N}{2M}$ for large $1<< N\le M$. Noting that
$\ln (N)$ is the maximal possible value of entropy of the subsystem $A$, 
the average entanglement entropy of a random pure
state was concluded to be
near maximal. Later, the same result was shown to hold for the 
$\beta=1$ case~\cite{Arul1}. On the other hand, there have been
only few studies on the full probability distribution of
the entanglement entropy. 
The distribution of the so called
G-concurrence~\cite{CSZ}, a measure of entanglement, was 
computed exactly in the
large $N$ limit and was shown to have a point measure (delta function).
For small $N$, the distribution
of purity is known~\cite{Giraud}. On the other hand, for large $N$,
the Laplace transform of the distribution of purity (for positive
Laplace variable) was computed in 
~\cite{Facchi1,Pasq} which only gave partial information about
the full purity distribution. 

Recently, using a Coulomb
gas approach, the full probability distribution of the
Renyi and von Neumann entropy, as well as that of purity, was 
computed exactly
in the large $N$ limit by studying the associated Coulomb gas
model via a saddle point method~\cite{NMV}. 
Interestingly, the
pdf of the entropy exhibits two singular points which correspond
to two interesting phase transitions in the Coulomb gas 
problem~\cite{NMV,Facchi1,Pasq}. Similar phase transitions
in the Coulomb gas picture, leading to a nonsingular pdf
of a physical observable, have also been noted recently in
several other problems where the random matrix theory is applicable:
these include the pdf of the conductance and the shot noise power
through a mesoscopic cavity~\cite{VMB1,VMB2,OK} (see also the chapter 35
and 36 of this book for applications of the RMT 
to quantum transport properties), the pdf of the number
of positive eigenvalues (the so called index) of Gaussian
random matrices~\cite{index}, nonintersecting Brownian
interfaces near a hard wall~\cite{Nadal} and in information
and communication systems~\cite{Kazak} to name a few.

Here our focus is on the statistical properties of the minimum
eigenvalue $\lambda_{\rm min}$ and for all values of $M=N$.
For the special case 
$\beta=2$
and $M=N$, the average value $\br \ldmin\kt$ 
was studied recently by Znidaric~\cite{Znd}. He computed, by hand, 
$\br \ldmin\kt$ for small values of $N$ and conjectured
that $\br \ldmin\kt=1/N^3$ for all $N$. Later, in~\cite{MBL}, 
the full
probability distribution of $\lambda_{\rm min}$ was computed explicity 
for all $M=N$ and $\beta=1$ and $\beta=2$. Znidaric's
conjecture for $\beta=2$ then followed as a simple corollary~\cite{MBL}.
In the next section, I briefly
outline this derivation and also provide a new result for
the distribution of $\lambda_{\rm min}$ for $\beta=4$.

\sect{Minimum Eigenvalue Distribution for $M=N$}\label{minimum}

In this section, we compute the distribution of $\lambda_{\rm min}$
when the eigenvalues are distributed via Eq. (\ref{jpdf1}).
It is easier to compute the cumulative distribution
\beq
R_N(x)=\mbox{Prob}\left[ \ldmin \ge x \right]=
\mbox{Prob}\left[ \ld_{1} \ge x, \ld_2 \ge x, \ldots, \ld_N \ge x \right].
\eeq
Using (\ref{jpdf1})
\beq
R_N(x)= B_{M,N}\, \int_x^{\infty}\cdots \int_x^{\infty}  \delta\left(\sum_{i=1}^N
\ld_i -1 \right)
\prod_{j<k} |\ld_j-\ld_k|^\beta\, \prod_{i=1}^N
\ld_i^{\f{\beta}{2}(M-N+1)-1}\, d\lambda_i.
\label{qmin}
\eeq
The challenge is to evaluate this multiple integral. 

We proceed
by introducing an auxiliary integral
\beq
I(x,t)= \int_x^{\infty}\cdots \int_x^{\infty}  \delta\left(\sum_{i=1}^N
\ld_i -t \right)
\prod_{j<k} |\ld_j-\ld_k|^\beta\, \prod_{i=1}^N
\ld_i^{\f{\beta}{2}(M-N+1)-1}\, d\lambda_i.
\label{aux1}
\eeq
If we can evaluate $I(x,t)$ for all $t$, then
\beq
R_N(x) = B_{M,N}\, I(x,1).
\label{qmin1}
\eeq
To evaluate $I(x,t)$, it is natural to consider its Laplace transform
\beq
\int_0^{\infty} I(x,t) e^{-st} dt = \int_{x}^{\infty} \cdots
\int_{x}^{\infty} e^{-s \sum_{i=1}^{N} \ld_i} \prod_{j<k} |\ld_j-\ld_k|^{\beta}\,
\prod_{i=1}^N \ld_i^{\f{\beta}{2}(M-N+1)-1}\, d\lambda_i.
\label{lt1}
\eeq
Next, a change of variable $\lambda_i= \frac{\beta}{2s}\, w_i$ reduces 
it to
\beq
\int_0^{\infty} I(x,t) e^{-st} dt = \left(\frac{\beta}{2s}\right)^{-\beta MN/2} 
\, \int_{2sx/\beta}^{\infty} \dots
\int_{2sx/\beta}^{\infty} e^{-\frac{\beta}{2}\sum_{i=1}^N w_i}\, 
\prod_{j<k} |w_j-w_k|^{\beta}\,
\prod_{i=1}^N w_i^{\alpha\beta/2}\, d w_i
\label{lt2}
\eeq
where $\alpha=(1+M-N)-2/\beta$.
Next we recognize the multiple integral, up to an overall constant, as the
cumulative distribution $Q_N(2sx/\beta)$ of the minimum eigenvalue $w_{\rm 
min}$ in the unconstrained Wishart ensemble discussed previously. Thus, up to an overall
constant $A_1$ independent of $s$, we have
\beq
\int_0^{\infty} I(x,t) e^{-st} dt= A_1 s^{-\beta MN/2}\, Q_N\left(\frac{2sx}{\beta}\right).
\label{lt3}
\eeq
The program then is to invert this Laplace transform, compute $I(x,t)$ for all $t$
and calculate $R_N(x)$ using Eq. (\ref{qmin1}). Henceforth, we will
drop the overall constant which can be finally fixed from the normalization
that $R_N(0)=1$.

Thus, if we know the cumulative distribution of the Wishart
minimum eigenvalue $Q_N(z)$, we can, at least in principle, determine
the minimum eigenvalue distribution $R_N(x)$ for the random pure state
problem. This is hardly surprising given the microcanonical to
canonical correspondence between the two ensembles. In practice, however,
it is nontrivial to invert the Laplace transform in Eq. (\ref{lt3}).
That indeed is the real challenge. We will see below that
fortunately for $M=N$, where $Q_N(z)$ is given in Eqs. (\ref{b1}), (\ref{b2})
and (\ref{b4}) for $\beta=1$, $2$ and $4$ respectively, 
this Laplace inversion can be carried out
in closed form and one can compute $R_N(x)$ 
explicitly in all three cases $\beta=1$, $2$ and $4$.\\

{\noindent \bf{The case $\beta=2$:}} Let us start with the simplest case $\beta=2$ with
$M=N$.
Here $Q_N(z)= e^{-Nz}$ from Eq. (\ref{b2}). Dropping the overall constant,
Eq. (\ref{lt3}) gives
\beq
\int_0^{\infty} I(x,t) e^{-st} dt = \df{e^{-sNx}}{s^{N^2}}.
\eeq
The Laplace inversion is trivial upon using the convolution theorem giving
(up to an overall constant) $I(x,t)= (t-Nx)^{N^2-1} \Theta(t-Nx)$
where $\Theta(z)$ is the step function. Putting $t=1$ and using (\ref{qmin1})
gives the exact distribution~\cite{MBL}
\beq
R_N(x)=\mbox{Prob}\left[ \ldmin \ge x \right] =\left(1-N x\right)^{N^2-1}
\Theta\left(1-Nx \right).
\label{udist}
\eeq
Subsequently, the pdf is given by
\beq
P_N(x)=-\frac{dR_N(x)}{dx}= N(N^2-1) (1-Nx)^{N^2-2}\,\Theta(1-Nx).
\label{updf}
\eeq
A plot of this pdf can be found in Fig. (\ref{fig:mindist}) for $N=2$. Thus $P_N(x)$ in 
$x\in [0,1/N]$ has the
limiting behavior
\begin{eqnarray}
P_N(x) & \to & N\,(N^2-1) \quad\quad {\rm as} \quad x\to 0 \nonumber \\
       &\to & N\,(N^2-1)\,(1-Nx)^{N^2-2} \quad {\rm as} \quad x\to 1/N
\label{clim}
\end{eqnarray}
One can easily compute all the moments explicitly
\beq
\mu_k(N)= \br \ldmin^k \kt=\int_0^{\infty} x^k P_N(x)\,
dx=\frac{\Gamma(k+1)\Gamma(N^2)}{N^k\, \Gamma(N^2+k)}.
\label{umoments}
\eeq
In particular, for $k=1$, we obtain for all $N$
\beq
\mu_1(N)=\br \ldmin \kt=\frac{1}{N^3},
\label{uavg}
\eeq
proving the conjecture by Znidaric~\cite{Znd}.
Putting $k=2$ in Eq. (\ref{umoments}), we get the second moment $\mu_2=\frac{2}{N^4
(N^2+1)}$.
Thus the variance is given by
\beq
\sigma^2= \mu_2(N)-[\mu_1(N)]^2= \frac{1}{N^6}\left(\frac{N^2-1}{N^2+1}\right).
\label{uvar}
\eeq
\begin{figure}[htb]
\begin{center}
\includegraphics[width=8cm]{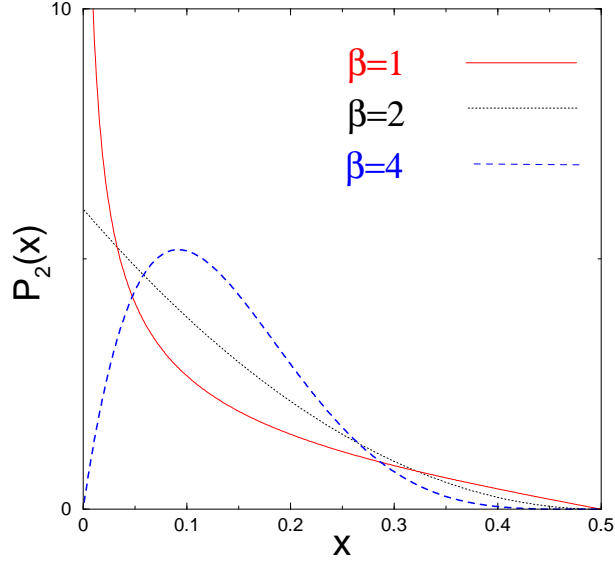}
\caption{The pdf of $\lambda_{\rm min}$ for 
$N=2$ for $\beta=1$, $\beta=2$ and $\beta=4$.
\label{fig:mindist}}
\end{center}
\end{figure}

{\noindent \bf{The case $\beta=1$:}} The computation in this case proceeds as in
the case $\beta=2$, though the Laplace inversion is nontrivial. Omitting
details~\cite{MBL}, we just quote here the main results.
For $M=N$, the pdf of the minimum eigenvalue, $P_N(x)= -dR_N(x)/dx $ 
is nonzero in $x\in [0,1/N]$ and is given by~\cite{MBL}
\beq
\label{resultreal}
P_N(x)=A_N\, x^{-N/2}\, (1-Nx)^{(N^2+N-4)/2} \, _2F_1\left(\f{N+2}{2},\f{N-1}{2},
\f{N^2+N-2}{2},-\f{1-Nx}{x} \right)
\eeq
where $_2F_1(a,bc,c,z)$ is the standard hypergeometric function~\cite{AS} and
the constant $A_N$ is given by
\beq
A_N=\df{N\, \Gamma(N)\, \Gamma(N^2/2)}{2^{N-1}\,\Gamma(N/2)\, \Gamma((N^2+N-2)/2)}.
\label{AN}
\eeq
The limiting behavior of $P_N(x)$ as $x\to 0$ and $x\to 1/N$ can be worked out
\begin{eqnarray}
P_N(x) &\approx & \left[\frac{\sqrt{\pi}\,
\Gamma(N)\,\Gamma(N^2/2)}{2^{N-1}\,\Gamma^2(N/2)\,\Gamma((N-1)/2)}\right]\, x^{-1/2}\quad\,
{\rm
as} \quad x\to 0 \nonumber \\
&\approx & A_N\, N^{-N/2}\, (1-Nx)^{(N^2+N-4)/2} \quad\, {\rm
as} \quad x\to 1/N
\label{rlim}
\end{eqnarray}

All moments can also be worked out explicitly~\cite{MBL}. In particular, the average
is given by
\beq
\mu_1(N)= \br \ldmin \kt = \frac{\sqrt{\pi}\,\Gamma(N)}{N\, \Gamma(N/2)\,\Gamma((N+5)/2)
2^{N-1}}\,
_2F_1\left(3,\frac{3}{2},\frac{N+5}{2},1-N\right).
\label{ravg}
\eeq
Thus the expression for $\br \ldmin \kt$ for arbitrary $N$ in the real $(\beta=1)$ case is
considerably more complicated than its counterpart in Eq. (\ref{uavg}) for the complex case.
One finds, from Eq. (\ref{ravg}), that $\mu_1(N)$ decreases with increasing $N$, e.g., 
$\mu_1(1)=1$,
$\mu_1(2)=(4-\pi)/8$, $\mu_1(3)=(2-\sqrt{3})/9$ etc. One can show~\cite{MBL} that
asymptotically for large $N$, $\mu_1(N)$ decays as
\beq
\mu_1(N) \approx \frac{c_1}{N^3}
\label{asympb1}
\eeq
where $c_1$ is precisely the constant in Eq. (\ref{c1}).\\

{\noindent {\bf {The case $\beta =4$:} } For $\beta=4$, we first substitute
$Q_N(z)$ from (\ref{b4}) in (\ref{lt3}), expand the hypergeometric function
in power series as in (\ref{dhyper}) and then invert the Laplace transform term by term
to get a series for $I(x,t)$.   
To transform each term, we make use of the convolution theorem. 
We then put $t=1$ in the expression for $I(x,t)$ and compute $R_N(x)$
in (\ref{qmin1}). The overall constant is fixed by imposing $R_N(0)=1$.
This gives the explicit expression, valid for $0\le x\le 1/N$, 
\begin{equation}
R_N(x) = \Gamma(N+1)\Gamma(2N^2)
\,\sum_{k=0}^N \frac{(2x)^k\, (1-Nx)^{2N^2-k-1}}{(N-k)!(2k)!\Gamma(2N^2-k)}
\label{b4cdist}
\end{equation}
The pdf $P_N(x)=-dR_N(x)/dx$ of $\lambda_{\rm min}$ vanishes linearly at $x=0$ as
$P_N(x)\to B_N\, x$ where $B_N= N(2N+1)(2N^2-1)(2N^2-2)/3$. This in contrast
to the $\beta=1$ case
(where $P_N(x)$ diverges as $x^{-1/2}$ as $x\to 0$) and also to the $\beta=2$ case where
$P_N(x)$ approaches a constant as $x\to 0$ (see Fig. (\ref{fig:mindist})).
At the upper edge, when $x\to 1/N$, the pdf vanishes as $P_N(x)\sim (1-Nx)^{2N^2-N-2}$.

All the moments can also be calculated explicitly for $\beta=4$. For example,
the average is given by
\begin{equation}
\mu_1(N)= \br \ldmin \kt =\frac{1}{2N^3}\sum_{k=0}^N {N \choose k}\, \frac{(k!)^2}{(2k)!}\, 
\left(\frac{2}{N}\right)^k.
\label{ravg4}
\end{equation}
One can extract the large $N$ asymptotics of this sum. We first 
express ${N\choose k} = \Gamma(N+1)/\Gamma(N-k+1)$, then use the property
of the Gamma function, $\lim_{z\to \infty} \Gamma(z+a)/\Gamma(z)\to z^a$,
to obtain for large $N$
\begin{equation}
\mu_1(N)= \br \ldmin \kt= \frac{1}{2N^3}\sum_{k=0}^{\infty} \frac{k!}{(2k)!}\, 2^k
\label{ravg4as}
\end{equation}
The sum can be exactly evaluated giving,
\beq
\mu_1(N)= \br \ldmin \kt
 \approx \frac{c_4}{N^3}
\label{asympb4}
\eeq
where $c_4$ is precisely the constant in Eq. (\ref{c4}).\\

Let us then summarize the behavior of the pdf $P_N(\lambda_{\rm min})$ of $\lambda_{\min}$ 
in the three
cases $\beta=1$, $\beta=2$ and $\beta=4$ (see Fig. (\ref{fig:mindist})).
At the lower edge $x\to 0$, $P_N(x)$ displays very different behavior in the three 
cases. As $x\to 0$, $P_N(x)$ diverges as $x^{-1/2}$ for $\beta=1$,
approaches a constant for $\beta=2$ and vanishes linearly for $\beta=4$.
On the other hand, at the upper edge $x\to 1/N$, $P_N(x)$ approaches
$0$ as a power law in all three cases, albeit with different powers, $P_N(x)\sim 
(1-Nx)^{\nu_\beta}$ where $\nu_{1}= (N^2+N-4)/2$, $\nu_2=N^2-2$ and $\nu_4=2N^2-N-2$.

In the large $N$ limit and in the range $x<<1/N$ (far away from the upper edge) the 
cumulative distribution $R_N(x)=\int_x^{1/N} P_N(x')\,dx'$
approaches the scaling form
$R_N(x) \to q_{\beta}(xN^3)$, where 
the scaling functions $q_{\beta}(y)$ are exactly same as 
in the unconstrained Wishart case given respectively in Eqs. (\ref{sc1}), (\ref{sc2})
and (\ref{sc4}). Thus, in this range and for large $N$, effectively 
the random variable $\lambda_{\rm min}$ in the bipartite problem behaves, in law, as
the Wishart minimum eigenvalue scaled by a factor $N^{-2}$, i.e.,
$\lambda_{\min} \to w_{\rm min}/N^2$.
This is also confirmed
in Eq. (\ref{avgtr1}), where we see that the average trace in the
unconstrained Wishart ensemble scales as $N^2$. Thus, in the microcanonical
enemble, where the trace is constrained to be unity, it amounts to
rescale all the Wishart eigenvalues by a factor $N^{-2}$ for large $N$.
However, for finite $N$, the distributions are very different in the
constrained and unconstrained ensembles, in particular near the upper edge
$x=1/N$. In other words, the distribution of $\ldmin$ exhibits
strong finite size effects.

\sect{Summary and Conclusion}\label{conclu}

In this chapter we have discussed an application of Wishart matrices
in an entangled random pure state of a bipartite system consisting
of two subsystems whose Hilbert spaces have dimensions $M$ and $N$
respectively with $N\le M$. 
The $N$ eigenvalues of the reduced density matrix of the smaller subsystem
are distributed exactly as the eigenvalues of a Wishart matrix, with the
only difference that the eigenvalues satisfy a global constraint: 
the trace is fixed to be unity. 

We have studied the distribution
of the minimum eigenvalue in this fixed-trace Wishart ensemble.
For the hard edge case (when two subsystems have same size $M=N$), we have shown
that the minimum eigenvalue distribution can be computed exactly for all $N$
in all three interesting physical cases $\beta=1$, $\beta=2$ and $\beta=4$.

What does this exact distribution of $\lambda_{\rm min}$ tell us
about the entanglement entropy of the bipartite system in a pure state?
We have seen before that if $\lambda_{\rm min}$ is close to its maximally allowed
value $1/N$ (set by the unit trace constraint), then that configuration
is maximally entangled since all the eigenvalues contribute equally to the
composition of the state. A measure of how close the random
state is to this maximally entangled state can be estimated by computing
the net measure of $\lambda_{\rm min}$ in a small range close to $1/N$, e.g.,
by the cumulative probability $R_N(1/N-\epsilon)=\int_{1/N-\epsilon}^{1/N} P_N(x)\,dx$
where $\epsilon<< 1/N$. From our exact calculation, we see that
the pdf of $\lambda_{\rm min}$, in all three cases, approaches zero
as $\lambda_{\rm min}$ approaches its maximum possible value $1/N$,
$P_N(\lambda_{\rm min}=x,N)\sim (1-Nx)^{\nu_\beta}$ where $\nu_1=(N^2+N-4)/2$, 
$\nu_2=N^2-2$ and $\nu_4= 2N^2-N-2$. This shows that the
`closeness to maximal entropy' measure $R_N(1/N-\epsilon)\sim (\epsilon\, 
N)^{\nu_\beta+1}$. For $\epsilon<<1/N$, this
measure is evidently very small.
It was argued before~\cite{Page95}, on
the basis of the computation of the only the first moment of the von Neumann entropy
(not the full distribution), that a random state is almost maximally entangled.
Our result shows that the {\em probability} that a random state is
maximally entangled is actually very small. The same conclusion
was also deduced recently on the basis of the large $N$ computation
of the full distribution of the Renyi entropy~\cite{NMV}.
Thus, the lesson is that conclusions based just on the first moment,
may sometimes be a bit misleading.

Here we have  discussed only the hard edge $M=N$ case
where the two subsystems have equal sizes. Our results are of relevance for small
systems such as when each subsystem consists of identical number of qubits.
It would be interesting to extend these calculations to the cases
when $M\ne N$. In particular,
in the context of thermodynamic systems where, for instance, one of
the subsystems is a heat bath, one needs to study the opposite limit 
$N<<M$. It would be
interesting to estimate the distribution of the minimum eigenvalue 
and other measures of entanglement 
in that limit. 

Finally, we have restricted ourselves here to `random pure' states where
all pure states are sampled equally likely.
This is the Haar measure. So far, we have not discussed dynamics, i.e.,
the temporal unitary evolution of the system. Under any unitary evolution
that is ergodic over the space of all pure states, Haar measure is
the unique stationary measure of the unitary evolution. 
This ergodicity holds
provided one does not have any strict conservation law.
For instance, the uniform measure {\em over all pure states} will not hold 
under the standard 
microcanonical scenario where the composite system has a fixed total energy
$E$ (eigenvalue of the Hamiltonian $\hat H$ of the composite system).
Only those pure states with total energy $E$, {\em and not all pure states}, 
will be 
sampled by the system under unitary evolution. An appropriate `stationary' 
measure is 
then the microcanonical measure which is uniform over all pure
states belonging to the fixed $E$ manifold.
For such a measure, one can again
define the reduced density matrix of the subsystem $A$ and its eigenvalues.
It would be interesting to study the statistics of the bipartite entanglement
(von Neumann or the Renyi entropy or the minimum eigenvalue $\ldmin$) 
in this microcanonical setting with a fixed total energy $E$. 
For such systems, some recent results
of very general nature (that do not require
the detailed knowledge
of the Hamiltonian of the system) 
have been derived~\cite{Popescu,Lebowitz,Reimann} which says that
any pure state, drawn from the uniform measure on the constrained
manifold, will almost surely be `maximally entangled' i.e., very close
to the maximally entangled (centroid) configuration $\lambda_i=1/N$ for all 
$i$, provided $M>>N$, i.e., the environment (subsystem $B$) is much bigger
than the system (subsystem $A$). It 
would be interesting to compute explicitly the distribution 
of this `typicality' i.e, the distance between the pure state
(drawn from a uniform measure over the constrained manifold)
and the maximally entangled state for some systems with specific
Hamiltonians.     

\vskip 0.3cm

{\sc Acknowledgements}: The discussion in this chapter is based on my joint 
work~\cite{MBL,NMV,VMB}
with O. Bohigas, A. Lakshminarayan, C. Nadal, M. Vergassola and P. Vivo. It is 
my pleasure to 
thank them. I am particularly grateful to C. Nadal for carefully
reading and correcting the manuscript.
I also thank my other collaborators on related subjects in random matrix 
theory: A. Comtet, 
D.S. Dean, 
A. Scardicchio, and G. Schehr. After this article was accepted, I came across
a recent preprint
by Y. Chen, D.-Z. Liu and D.-S. Zhou (arXiv: 1002.3975) 
where
the distribution of $\lambda_{\rm min}$ was computed for $M>N$ and for all
$\beta$. Also, the average density of eigenvalues for all finite $N$ and $M$
but for $\beta=1$ was computed in a recent preprint of P. Vivo.

\end{document}